\documentclass{epl}
\usepackage{graphicx} 
\usepackage{psfrag}
\usepackage{fancyhdr}  

\title{Why charges go to the surface: a generalized Thomson problem}
\shorttitle{Generalized Thomson Problem}
\author{Yan Levin\thanks{E-mail: \email{levin@if.ufrgs.br}} 
   \and Jeferson J. Arenzon\thanks{E-mail: \email{arenzon@if.ufrgs.br}}}
\institute{Instituto de F\'{\i}sica, 
  Universidade Federal do Rio Grande do Sul \\ 
  Caixa Postal 15051, CEP 91501-970, Porto Alegre, RS, Brazil}         

\pacs{71.10.C}{Electron gas-theories and models}
\pacs{64.60.Cn}{Order-isorder transformations; statistical mechanics of
model systems}

\date{\today}

\begin{document}    
\maketitle

\begin{abstract}
We study a variant of the generalized  
Thomson problem in which $n$ particles 
are confined to 
a neutral sphere and interacting by a $1/r^{\gamma}$ potential. 
It is found that for 
$\gamma \le 1$ the electrostatic repulsion expels all the charges to the 
surface of the sphere. However for $\gamma>1$ and $n>n_c(\gamma)$ occupation 
of the bulk becomes 
energetically favorable. It is curious to note that the Coulomb law lies 
exactly on the interface between these two regimes. 
\end{abstract}

In a recent paper~\cite{BoCaNe02} Bowick et al. studied 
a system of particles confined to the surface of a
sphere and interacting by a repulsive $1/r^\gamma$ potential
with $0<\gamma<2$. They called this ``the generalized Thomson
problem''. 
It is interesting, however, to recall that the original
Thomson problem was posed as a model 
of a classical atom~\cite{Th04}.   Thus, $n$ electrons 
were supposed to be confined in the
{\it interior} of a sphere with a uniform 
neutralizing background, the so called
``plum pudding'' model of an atom. 
The Thomson problem, which is still unsolved, is then  to
find the ground state of electrons {\it inside} the sphere.

In the absence of a  neutralizing background
the electrostatic repulsion between the  particles
``dynamically'' drives the charges to the surface. 
This significantly simplifies the calculations by reducing the search of
the ground state from the three dimensions down to two~\cite{SaKu97}.
But what
if instead of  the Coulomb potential  electrons interacted by
a  $1/r^\gamma$ potential?  Would they still
go to the surface or prefer to stay in the bulk? This
question was not addressed in the paper of
Bowick et al. who have {\it a priory} confined their particles
to reside on the surface.

It is clear that for a small number of charges, mutual repulsion will
force them to the surface.  What happens, however, 
as the concentration of particles increases?  To answer this question
we  compare the electrostatic energy of the configuration in 
which all $n$ particles are on the surface of a sphere with a 
configuration in which $n-1$ particles are at the surface and
one particle is located at the center of a sphere. 
The electrostatic energy of $n$ particles of charge $q$ interacting
through
a generalized Coulomb potential $q^2/\epsilon r^\gamma$, with dielectric
constant $\epsilon$, confined to the
surface of a sphere with  radius $a$ 
can be obtained by considering the electrostatic energy 
of the two dimensional one component plasma (OCP) $F_n^{OCP}$, {\it i.e.}
charges on the surface of a sphere with a neutralizing 
background ~\cite{Le02,Sh99a,MeHoKr01,AlAmLo98},
\begin{equation}
\label{0a} 
F_n^{OCP}=E_n+
\frac{q^2}{2 \epsilon a^\gamma} \frac{2^{1-\gamma}}{2-\gamma} n^2-
\frac{q^2}{\epsilon a^\gamma} \frac{2^{1-\gamma}}{2-\gamma} n^2 \;.
\end{equation}
The first term $E_n$ is the electrostatic energy of mutual repulsion
between the charges, the second term is the self energy of the neutralizing
background, and the third term is the energy of interaction between
the charges and the background.  The advantage of working with 
the one component plasma is that its ground state energy can be 
estimated by considering the interaction of an individual charge with the
background inside its Wigner-Seitz cell.  The characteristic
distance $d$ between the charges on the sphere is such that,
$\pi d^2 n =4 \pi a^2$, and
\begin{equation}
\label{0b} 
d=\frac{2 a}{\sqrt n}\;.
\end{equation}
The ground state energy of a Wigner crystal of charges interacting by
a $1/r^\gamma$ potential is, therefore,
\begin{equation}
\label{0c} 
F_n^{OCP}=-M_\gamma n\frac{q^2}{\epsilon d^\gamma}\;,
\end{equation}
where $M_\gamma$ is the Madelung constant.  
Of course, there is no perfect crystalline order of charges 
on the surface of the sphere and some topological defects 
must be present~\cite{BaBoCa03}.
Nevertheless, we expect that the topological defects will not
modify the scaling form of the Eq.~(\ref{0c}) but will only affect
the numerical value of the Madelung constant. 
Equation (\ref{0a}) can now be rewritten as
\begin{equation}
\label{1} 
E_n=\frac{q^2}{2 \epsilon a^\gamma}\left[\frac{2^{1-\gamma}}{2-\gamma} n^2-
\frac{M_\gamma}{2^{\gamma-1 }} n^{1+\frac{\gamma}{2}}\right]\;.
\end{equation}
This is equivalent to Eq. (4) of the reference~\cite{BoCaNe02}.   
For the case
of Coulomb interaction, $\gamma=1$, it has been known for 
some time~\cite{ErHo91}
that Eq.~(\ref{1}) with $M_1 \approx 1.102$ gives an excellent 
approximation to the ground state energy of electrons on the surface of
a sphere. This value of $M_1$ is, indeed, 
very close to the Madelung constant of a
planar Wigner crystal which is $1.1061$. 
To test the accuracy of Eq.~(\ref{1}) for different values of $\gamma$ 
we have simulated the distribution
of charges on the surface of a sphere and calculated their electrostatic
energy.  To speed up the simulations, instead of using the full continuum
algorithm for the surface Thomson problem, we have studied its 
discrete version.  Thus, a  number of sites were randomly placed
on the surface of the sphere with a uniform probability density.  
The charges were then restricted to move only on these sites.  There
is a significant gain in the simulation time since all the electrostatic 
interactions can be tabulated once at the beginning of the simulation.

In figure 1 we compare the result of numerical simulation with  
the analytic expression given by Eq.~(\ref{1}). The agreement is perfect
for the whole range of $n$.  
\begin{figure}[h]
\begin{center}
\includegraphics[angle=270,width=10cm]{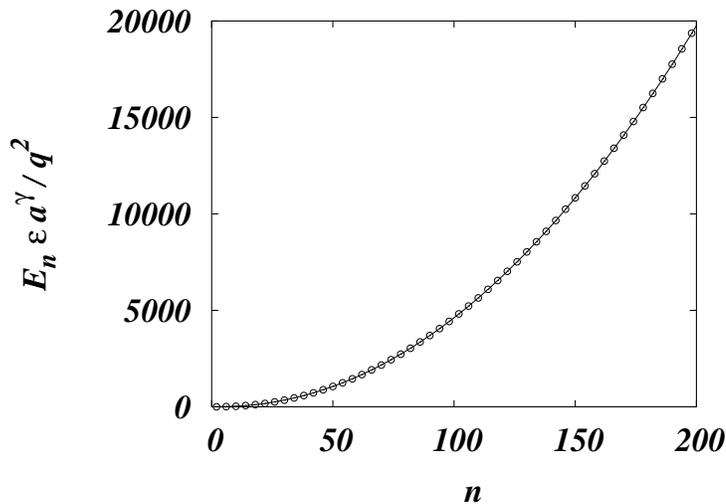}
\end{center}
\caption{Energy $E_n$ for charges with $\gamma=1.4$, 
Eq.~\ref{1}, as a function of the
number $n$ of particles, compared with the simulation
data (points). The Madelung constant
$M_{\gamma}\simeq 1.78$ gives a perfect fit to the data points
over the whole range on $n$. Were used $2000$ sites for 
allowed positions of the charges on the surface of the sphere.}
\label{figure1}
\end{figure}

Since the number of metastable states grows
exponentially~\cite{ErHo95} with increase in $n$, 
our simulation is not able to locate the exact ground
state for large number of charges.  However,
the energy of nearly degenerate metastable states is very close
to that of the ground state, and the  error thus accrued
is minimal. This can be clearly seen from the  absence
of any visible fluctuations in the data points plotted in Fig. 1.

We are now in position
to compare the electrostatic energy of the configuration in which 
all the  electrons are at the surface, with the configuration in which 
$n-1$ particles are
on the surface and one charge is at the center of the sphere,
\begin{equation}
\label{2} 
\Delta E(n)=E_{n-1}+\frac{q^2 (n-1)}{\epsilon a^\gamma}-E_n\;.
\end{equation}

\begin{figure}[h]
\begin{center}
\includegraphics[angle=270,width=10cm]{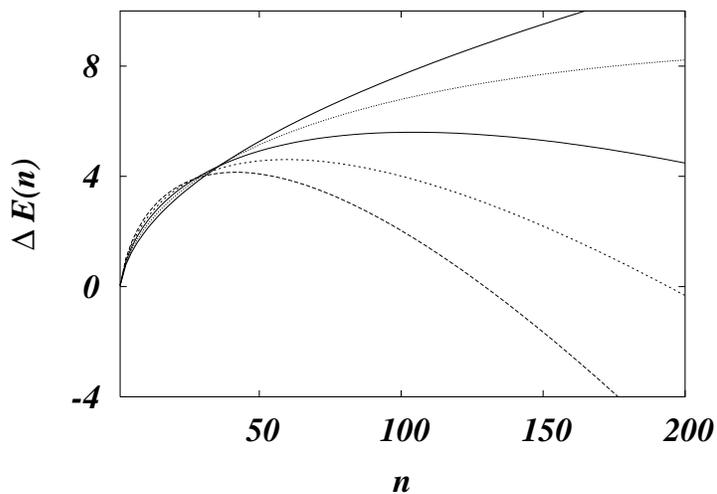}
\end{center}
\caption{The difference in energy $\Delta E(n)$ 
between the configurations in which
one particle is at the center of a sphere and $n-1$ particles are at the
surface, and the configuration in which
all $n$ charges are on the surface. The values of $\gamma$ range from $1$ for
the topmost curve, to $1.4$ for the bottommost curve, in intervals of $0.1$. 
The topmost curve $\gamma=1$ is a monotonically increasing function of $n$, 
while for $\gamma>1$ the curves after reaching a maximum decline.}
\label{figur2}
\end{figure}

From Fig. 2 we see that for $\gamma > 1$,  $\Delta E(n)$ 
starts positive, so that the  charges are driven to the surface.
However, as the surface particle population increases, $\Delta E(n)$ reaches
a maximum and begins to declines. At the threshold number of charges 
$n=n_c(\gamma)$ it becomes
energetically  favorable for the particles to penetrate into the bulk.  
For $\gamma \rightarrow 1^+$ the critical number of charges diverges as,
\begin{equation}
\label{3} 
n_c \sim \frac{1}{(\gamma -1)^2}\;.
\end{equation}
It is very curious that the Coulomb law is precisely
at the border line of the two regimes.  For the potentials with
$\gamma \leq 1$ repulsion is strong enough to drive all the charges to the
surface, while for  $\gamma > 1$ and $n>n_c(\gamma)$ the charges will
penetrate into the bulk.
Unlike for Coulomb charges, for particles interacting by $1/r^\gamma$ 
potential, the {\it Thomson problem} 
requires a full three dimensional analysis of the distribution of 
particles inside the sphere. The {\it generalized} Thomson problem is, 
therefore, much more complex than its classical counterpart.

\stars
This work was supported in part by the Brazilian agencies
CNPq and FAPERGS.


\end{document}